\documentclass{ptptex}

\usepackage{graphicx}

\title{
Discrimination of Quark Stars from Neutron Stars\\
in  Quadrupole Oscillations%
}
\author{
Yasufumi \textsc{Kojima} 
and
Koh-ichi \textsc{Sakata}%
}

\inst{
Department of Physics,
Hiroshima University,
Higashi-Hiroshima 739-8526
Japan
}

\abst{%
The frequencies and damping times due to gravitational radiation
are calculated for self-bound quark star models. The results are
compared with those for neutron star models.
They are markedly contrasted in less
relativistic cases. 
The distinction derived here from
a simple model of quark stars 
may be relevant to the future
theoretical and observational studies,
since the oscillation properties essentially 
depend on the mass and radius of an equilibrium star.
}

\begin{document}

\maketitle

\section{Introduction}

  Recently,  Chandra X-ray Observations reveals
a new aspect of a compact star. 
The isolated neutron star candidate RXJ1856.5-3754, 
which may be the closest to us, 
has the radiation radius 3.8-8.2 km  \cite{det}. 
The distance is estimated as  60-130pc
from the parallax  \cite{wal}\cite{kka}  
and the interstellar column density \cite{det}. 
The ambiguity leads to the range of the estimated
size, but the value is evidently much smaller
than the canonical radius of a neutron star $\sim 10$ km
\cite{lp}.

The radiation radius $ R_\infty$
is determined by the observed energy flux and
blackbody temperature.
Assuming spherical symmetric equilibrium,
$ R_\infty$ is expressed by the radius 
$R$ in the Schwarzshild coordinate as 
$R_\infty = R(1-2GM/Rc^2)^{-1/2}$,
where $M$ is the gravitational mass.
It is clear that  $ R \le R_\infty $.
After some algebra, we also have
$ GM/c^2 \le R_\infty /(3\sqrt{3}) . $ 
The gravitational mass must be less than
 0.49-1.05 $M_\odot$
corresponding to the observed value of $ R_\infty$.
This limit of mass is significantly less than the 
canonical mass of a neutron star $ \sim 1.4 M_\odot $
\cite{lp}.
In this way, the characteristic size and mass are
unusual for any neutron star models.

There are however several attempts 
to reconcile with the standard neutron star models. 
For example, the atmosphere is metal dominated one, 
and/or the surface temperature distribution is inhomogeneous.
At present, there is no positive evidence to support them
even in the long term and high-resolution observation \cite{det}.

Taken at face value, 
RXJ1856.5-3754 is not a neutron star, but
a different species of a compact star, 
which is likely to be  a quark star.
We would definitely judge the state of the compact star,
if there is different information such as mass, which 
is not easily determined by the single star.
Nakamura \cite{nak} proposed a possible 
formation scenario of the quark star.
In the standard scenario to neutron star after supernova,
it is difficult to produce such light remnant 
$\sim 0.7 M_\odot$ 
leading to the quark star. If the progenitor has 
significant rotation, small core is left
through the centrifugal breakup.
In that case, huge amount of energy can  be radiated 
by gravitational waves.
The quadrupole oscillation is likely to be driven by such 
a collapse, and the asymmetry of matter distribution is 
efficiently smoothed out by the gravitational radiation.
The detection of such gravitational waves
significantly depends on  the number of the events. 
There is a big uncertainty about it, because we
know just one candidate, RXJ1856.5-3754.

Motivating the observational suggestion,
we will study the property of the quark star models.
Our concern is not the equilibrium structure, which
was already calculated in the literature
\cite{lp}\cite{afo}\cite{hzs}\cite{gle}~, 
but the dynamical aspect,
that is, non-radial oscillations resulting from 
small disturbances. In particular, we compute
the characteristic frequency of
the quadrupole f-mode oscillations, 
which is the most important for the gravitational radiation.
Such calculations were extensively 
performed for several non-rotating 
neutron stars 
\cite{lindet}\cite{cls}  and
slowly rotating polytropic models
\cite{koj92}\cite{koj93apj} .
By extending to the quark star models,
the 'catalog' is enriched
for the future gravitational wave astronomy.
In addition to direct gravitational wave signal,
oscillatory phenomena after some kinds of 
sudden bursts or flares 
may appear in X-ray and/or gamma-ray observation.
The global properties such as mass and radius
should be relevant to dynamical overall oscillations
on the star.
Non-spherical distortion efficiently 
decays due to gravitational wave emission
within a few second as shown later. 
The typical oscillation frequency is kHz.
These properties, 
irrespective of observational bands or tools,
provide useful diagnosis for the interior of 
compact stars.
Other variability except f-mode 
may also appear in different frequency range, 
but they are much reflected by details of
the interior structure, which are not so clear
at moment. Therefore, we do not consider them here. 
Microscopic equation of state for high-density matter 
is not yet established in neutron stars and quark stars.
We use a simplest one, which may be helpful within 
our present knowledge of high-density matter
to understand the contrasts
between neutron stars and quark stars.
The model and the numerical method are
summarized in \S 2.  The results are shown in \S 3.
In \S 4, we compare the self-bound quark stars
with neutron stars in  their oscillation 
frequencies and the decay times.
Finally in \S 4, concluding remarks are given.

\section{Models and Calculations}

Witten \cite{wit} 
conjectured that three-flavor quark matter consisting 
of u, d, and s quarks may be actually be the ground 
state of the strong interaction.
If this strange quark matter hypothesis is true,
we expect pure quark matter star.
The properties of the strange star are
explored using a simple microphysical treatment of
quarks \cite{afo}\cite{hzs}\cite{gle}~.
The quark matter is described by 
the fundamental theory of the strong interaction, QCD.
The equation of state (EOS) for it can be derived 
in principle.
However, there are not yet practical results
due to the difficulty of the theory.
Most important aspects of the strong interaction are
asymptotic freedom and confinement of quarks. 
They are described by phenomenological treatment,
the MIT bag model, i.e., 
free massless quarks are confined within the bag.
Based on such a simple picture,
the EOS can be written by a sum of the pressure by free quarks
and the bag pressure $-B$ \cite{gle} .
\begin{equation} 
 p = \frac{1}{3} \rho c^2 -\frac{4}{3} \frac{B}{(\hbar c)^3} .
\label{eos}
\end{equation} 
%

The equilibrium stellar models can be 
constructed by solving the TOV equation. 
The self-bound star has a sharp edge at the surface,
corresponding to $p =0$, and therefore
$ \rho c^2 =4 B/(\hbar c)^3 $.
The density is high enough, and
the interior density distribution does not so drastically  
change in the magnitude from the center to the surface.
The density distribution of self-bound stars and the 
comparison with neutron stars are fully explained 
elsewhere \cite{gle} .
For the simple EOS
(\ref{eos}),
the equilibrium models have a scaling law with
the bag constant $B$.
In terms of the radii and masses of one sequence of
stars with $B$, those for any other choice of $B'$
can be found from 
\begin{equation} 
R(B') =\sqrt{B/B'} ~R(B),
~~~~
M(B') =\sqrt{B/B'} ~M(B).
\end{equation} 
%

In this paper, we consider the spherical symmetric
equilibrium models and the perturbations from them.  
The oscillations with small amplitude 
are described by the perturbation of the Einstein equation.
By the spherical symmetry of the background model,
the perturbation quantities may be decomposed into
spherical harmonics. We consider 
the f-mode of quadrupole $l=2$ only, which 
is the most important for gravitational radiation.
The properties of such oscillation were extensively
studied so far for non-rotating neutron stars constructed from
various EOS
\cite{lindet}\cite{cls}~.
The eigenfrequencies are determined by
solving a fourth-order system of  differential equations 
inside the star, and a second-order one outside the star.
They are subject to appropriate boundary conditions.
By imposing outgoing wave condition, 
the normal frequency becomes a complex number, that is, 
the real part $\sigma_R$ represents the oscillation frequency
and the imaginary part  $\sigma_I$ the damping rate.
The equations and numerical methods adopted here
are summarized in details elsewhere \cite{koj97ptp}.

\section{Results}

The oscillation frequency and the damping rate for
the EOS (\ref{eos}) were previously
calculated only for a few stellar models \cite{cls}.
We have here extended the similar kind of calculations to much 
larger range of stellar models
in order to understand the general properties. 
The angular frequency 
$\sigma_R$ is shown as a function of the compactness $GM/Rc^2$
in Fig.1.
The frequency is normalized by
$
\Omega_K = \sqrt{GM/R^3} .
$
In the limit of Newtonian case $GM/Rc^2 \to 0$,
the frequency of stellar model with (\ref{eos}) 
is well approximated by
$\sigma _f =\sqrt{2l(l-1)/(2l+1)}\Omega_K 
\simeq 0.894\Omega_K $
for $l=2$,
which is the frequency of the Kelvin f-mode
for the uniform density.
The frequency may be approximated by
that for the uniform density case, since
the density distribution is almost constant
in the self-bound quark star models.
The similarity to the constant density case
is much clear by comparing with polytropic EOS,
$ p=K \rho ^{1+1/n}$.
The results for $ n=$1, 0.5 and 0.2
\footnote{
A little bit careful treatment is necessary for
the limit $n=0$ 
in the relativistic polytropic models.
The pressure and density perturbations are related as
$ \delta p = C^2 \delta \rho $ in general, but
$C^2 \to \infty $ 
and  $\delta \rho  \to 0$ for $n \to 0$.
It is possible to calculate for 
$ n=0$ by modifying the relevant parts, 
but  such a task would lead to only a tiny difference.
}
are shown in Fig.1. 
The oscillation frequencies for the EOS (\ref{eos}) 
may be approximated like those of
stellar models with smaller $ n \approx 0.$
We have also calculated for smaller value of $n$.
The agreement in the Newtonian limit becomes
better as $n \to 0$, but
the overall agreement in some range of $GM/Rc^2 $
is impossible for any value of $n$.
The treatment like $n \approx 0$ in the polytropic model
is not exact, but just approximate one, as there is no reason
to support.

\begin{figure}[h]
\centering
\begin{minipage}{0.45\linewidth}
\includegraphics[scale=1.5]{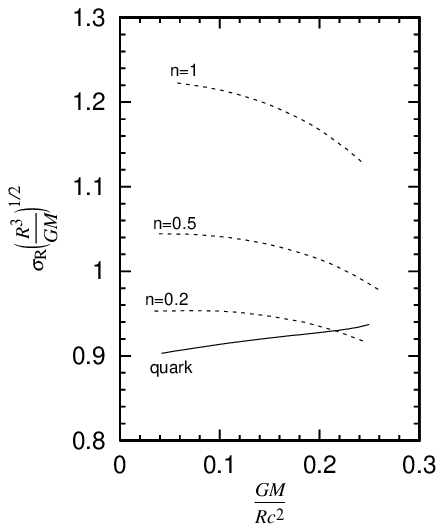}
\caption{
Angular frequencies of quadrupole f-mode
oscillation as a function of star compactness.
The solid and dotted lines correspond to
the quark and polytropic stars.
}
\end{minipage}
\hspace{8mm}
%
\begin{minipage}{0.45\linewidth}
\includegraphics[scale=1.5]{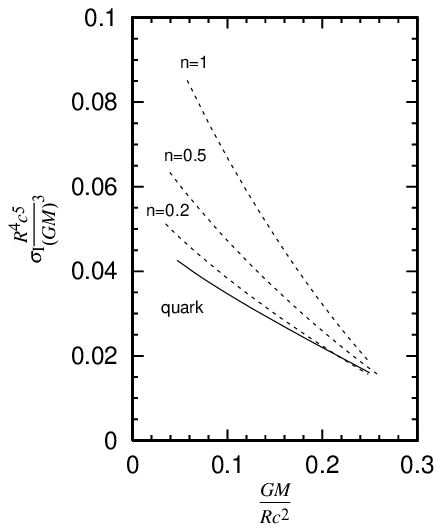}
\caption{
Decay rates of quadrupole f-mode
oscillation as a function of star compactness.
The solid and dotted lines correspond to
the quark and polytropic stars.
}
\end{minipage}
\end{figure}

In Fig.2, we show the damping rate $\sigma_I$  of the mode.
The characteristic decay rate can be evaluated by
the energy loss rate by the quadrupole radiation formula,
$ \sigma_I = dE/dt/(2E) \propto (GM)^3/(R^4 c^5).$  
We use this scale as the normalization in Fig.2.
The decay rate of the Newtonian star with uniform density
is analytically calculated as
$\sigma _I (R^4 c^5)/(G^3 M^3)= 32/625
= 5.12 \times 10^{-2} .$
The decay rate  of the  star with (\ref{eos})
coincides with this value in the Newtonian limit, as
the oscillation frequency does.
The dependence of the compactness
is also very similar to the model with small $n$.
In this way, the general behavior 
of the frequency and the decay rate 
for the EOS (\ref{eos}) can be described 
by the results for the stellar models 
with small $ n \approx 0.$

\section{Comparison among various models}

The frequency of self-bound quark star model
also has the scaling,
$ \sigma \propto \sqrt{M/R^3} \propto \sqrt{B} .$
The bag constant $ B$ is not exactly determined,
but is usually chosen as
$ B^{1/4} =$ 145-165 MeV
\cite{lp}\cite{afo}\cite{hzs}\cite{gle}~.
The approximate form (\ref{eos}) 
itself would be meaningless by theoretically 
considering realistic effects such as the quark mass
and gluon interaction.
We use $ B^{1/4} = 155$ MeV,
i.e., the energy density 
$ B /(\hbar c)^{3}  = 75 $MeV/fm$^3$
as a representative example.
The energy density at the stellar surface 
$4B/(\hbar c)^{3}$ corresponds to
about twice the energy of normal
nuclear matter $\rho_n c^2 =141 $MeV/fm$^3$,
i.e.
$\rho_n =2.5 \times 10^{14}$ g/cm$^3$.
We also show the results scaled to
$  B^{1/4} = 180$ MeV,
i.e.,  $ B  /(\hbar c)^{3}= 137 $MeV/fm$^3$.
In Fig.3, we show the frequency  $\nu =\sigma _R/2\pi$
of quadrupole oscillation along
the sequence of quark stars with a fixed value of $B$.
We also include the results for the
various neutron star models \cite{lindet}~.
A lot of EOSs are proposed for the high-density neutron matter,
but only selected EOSs are used.
Our collection might be old-fashioned, but
covers from soft to stiff EOSs. 
We omitted too soft ones, which cause the
maximum mass of neutron star below 1.4 $M_\odot $. 
The four models are denoted by
A \cite{ar72}~,  BJ \cite{bj74}~, 
PPS-T and  PPS-M \cite{pps76}
in Fig.3.
The stiffness is roughly in this order,
A is the softest and PPS-M the stiffest.
The description of these EOSs is also written elsewhere 
\cite{lindet}~.
%

There is quite different dependence on the compactness 
$GM/Rc^2 $ 
in the frequencies of gravitationally bound 
and self-bound stars.
In the equilibrium sequence of neutron stars,
the oscillation frequency increases with
$GM/Rc^2 ,$
because the mass increases  and
the radius decreases,  and hence
$\nu \propto \sqrt{GM/R^3} $ strongly increases. 
In order to understand this property, 
we labeled the average densities
$3M/4\pi R^3$ along each equilibrium sequence
in Fig.3. 
They are normalized by the nuclear density $\rho_n$.
On the other hand, 
both mass and radius in the sequence of self-bound quark star 
models increase as $ M \propto R^3 , $
and hence  the oscillation frequency is almost constant,
i.e.,
$\nu \propto \sqrt{GM/R^3} \approx $ constant. 
Actually the average density of 
the model  $ B^{1/4} = 155$ MeV,
changes within 2.1-3.2 $\rho_n$,
and that of $ B^{1/4} = 180$ MeV 
within 3.9-5.9 $\rho_n$.
The frequency $\nu$ of a quark star is approximated as
\begin{equation}
 \nu  = \frac{\sigma_R}{2\pi} \approx 1.7 \times
\left\{
1+4 \left( \frac{GM}{Rc^2} \right)^2
\right\}
\left(\frac{B^{1/4}}{155{\rm MeV}} \right)^2
{\rm kHz}.
\end{equation}
The dependence on the bag constant $B $ is exact, but 
relativistic correction is introduced to be fitted
for $GM/Rc^2 \le 0.25$ in the  braces.
In general, the frequencies of quark stars  
are discriminated from those of neutron stars
in less relativistic regime, $GM/Rc^2 \approx 0.$
In that case,  
the radius of the quark star  is much smaller, and the 
frequency is higher.

In Fig.4,  we show the decay time $\tau =1/ \sigma_I$
for self-bound quark star and neutron  stars.
The decay times of both stellar models strongly 
depend on the compactness $GM/Rc^2 .$
There is not so much clear difference 
unlike the oscillation frequency.
The decay time $\tau $ for the EOS (\ref{eos}) 
is approximated as
\begin{equation}
 \tau  = \frac{1}{\sigma_I} \approx 1.7 \times 10^{-3}
\left\{
1+12 \left( \frac{GM}{Rc^2} \right)^{3/2}
\right\}
\left(\frac{GM}{Rc^2} \right)^{-5/2}
\left(\frac{B^{1/4}}{155{\rm MeV}} \right)^{-2}
{\rm s}.
\end{equation}
The formula provides exact dependence on the bag constant $B $ and
the value in the Newtonian limit. The
expression in the braces is the relativistic correction 
to be fitted for $GM/Rc^2 \le 0.25.$ 
In Figs.3 and 4, we only show the results for
two choices of $B$, 
but one can easily calculate for the other choices 
by eqs. (3) and (4).
%

\begin{figure}[h]
\centering
\begin{minipage}{0.45\linewidth}
\includegraphics[scale=1.5]{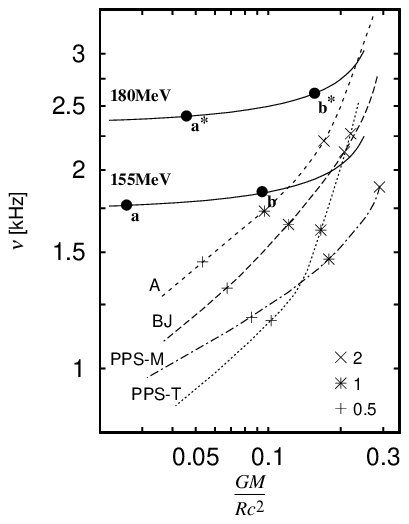}
\caption{
Frequencies in units of kHz 
as a function of star compactness.
The solid lines correspond to
the quark star models with
two different values of the bag constant.
Others denote neutrons star models
based on different equations of state.
The symbols on the curves of neutrons star models
mean the average density normalized by the nuclear density. 
See the text for details.
}
\end{minipage}
\hspace{10mm}
%
\begin{minipage}{0.45\linewidth}
\includegraphics[scale=1.5]{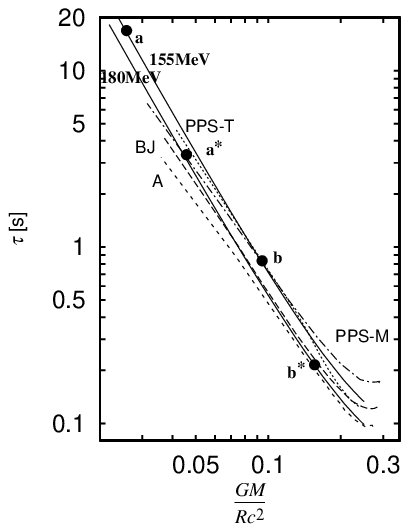}
\caption{
Decay times in units of second
as a function of star compactness.
The solid lines correspond to
the quark star models with
two different values of the bag constant.
Others denote neutrons star models
based on different equations of state.
See the text for details.
}
\end{minipage}
\end{figure}

We now apply our results to the parameters inferred from
the quark star candidate RXJ1856.5-3754.
The compact star has small radiation radius,  
3.8-8.2 km as mentioned in \S 1.
This information provides one constraint 
between the mass $M$ and radius $R.$
Assuming the canonical value 
$ B^{1/4} = 155$ MeV,
we have 
$ M= 0.068 M_\odot $,
$R= 3.8 $km, for  $R_\infty = $ 4.0km
and
$ M= 0.46 M_\odot $,
$R= 7.2 $km, for  $R_\infty =$ 8.0km.
Corresponding frequencies 
and decay times are marked as
'$a$' and '$b$' in Figs.3 and 4.
Expected frequency and decay time should
be within these points.
These values are shifted to
$ M= 0.12 M_\odot $, $R=3.8 $km, for  $R_\infty =$ 4.0km,
and $ M= 0.70 M_\odot $,
and $R= 6.6 $km, for  $R_\infty = $ 8.0km,
if $  B^{1/4} = 180$ MeV.
These points are labeled as
'$a^*$' and '$b^*$' in Figs.3 and 4.
From these figures, we can 
discriminate the quark stars from neutron stars
by the frequency and decay time of their oscillations.
%
Suppose that we have the information about
$\nu = \sigma_R/2 \pi$ and $\tau =1/\sigma_I$ 
at the same time.
As for a definite example, we assume   
the frequency $\nu \approx $ 1.7 kHz.
Since the oscillation frequency determines
the bag constant $B$ in the self-bound quark star model, 
we have $B^{1/4} \approx 155$ MeV.
Alternatively some neutron star models  with $GM/Rc^2  \geq 0.1 $
are also possible from Fig.3.
In other words, 
the average density is inferred as 1-2 $\rho_n$.
The decay times are quite different as seen from Fig.4. 
The decay times corresponding to the neutron stars
are 0.1-0.8 s, while
the longer range such as  1-10 s is possible for
the quark stars.
This is a speculation of the possibility and
a detailed argument
requires further study.

\section{Concluding remarks}

In this paper, we have presented suggestive 
results for the pulsations  
associated with gravitational radiation.
The gravitational waves carry key information
of the sources. We can in principle use them
as probes of compact stars.
The oscillation frequency scales with the average density
of a star $\nu \propto \sqrt{ M/R^3}$, 
since the f-mode has no radial node.
The decay time due to gravitational radiation 
significantly depends on the relativistic factor $ M/R.$
We may utilize these properties to discriminate
self-bound quark stars from gravitationally bound neutron stars.
If self-bound quark stars are realized in 
less relativistic regime, the oscillation frequency
is high enough due to high density, but decay time
is longer than that of neutron star.
In that case,
the mass and size are quite different
in the equilibrium sequences of two compact stars. 
It is clear that present self-bound
quark model is  too crude like the bag model.
The model is used to calculate explicitly,
but the resultant properties of the oscillations 
roughly depend on the 
macroscopic  values such as the mass and radius.
We therefore expect that such
discriminated properties 
should be involved in any detailed models.


Finally, we will comment on the astrophysical relevance 
of the quadrupole f-mode oscillation.
The driving mechanism like Cepheid variables
is not yet known, so that 
the persistent oscillations would not be observed. 
Rather, abrupt changes of the structure
like the formation\cite{nak}
and/or 
some kinds of bursts may be relevant. 
The origin and the nature of them are also unclear at moment.
However, it is evident that the gravitational wave emission 
is crucial in such dynamical process.
The most efficient mode is the quadrupole ($l=2$) f-mode,
which was calculated here.
We should await for the fully relativistic numerical 
calculation to simulate the violent events,
but the oscillation frequency 
may be used even for checking the constructing numerical 
code \cite{shb}.
%
\section*{Acknowledgements}
This work was supported in part
by the Grant-in-Aid for Scientific Research 
(No.14047215) from 
the Japanese Ministry of Education, Culture, Sports,
Science and Technology.


\end{document}